\documentclass{article}

  \usepackage[final]{neurips_2025_ml4ps}

\usepackage[utf8]{inputenc} 
\usepackage[T1]{fontenc}    
\usepackage{hyperref}       
\usepackage{url}            
\usepackage{booktabs}       
\usepackage{amsfonts}       
\usepackage{nicefrac}       
\usepackage{microtype}      
\usepackage{xcolor}         
\usepackage{float} 

\usepackage{tikz}
\usetikzlibrary{arrows.meta, positioning, calc, shapes.geometric}

\title{Exoplanet formation inference using conditional invertible neural networks}

\author{%
  Remo Burn \\
  Max Planck Institute for Astronomy\\
  Heidelberg, Germany \\
  \texttt{remo.burn@oca.eu} \\
  \And
  Victor F. Ksoll \\
  Institut f\"ur Theoretische Astrophysik \\
  Heidelberg, Germany\\
  \texttt{v.ksoll@uni-heidelberg.de}
  \And
  Hubert Klahr \\
  Max Planck Institute for Astronomy\\
  Heidelberg, Germany \\
  \texttt{klahr@mpia.de} \\
  \And
  Thomas Henning \\
  Max Planck Institute for Astronomy\\
  Heidelberg, Germany \\
  \texttt{henning@mpia.de} \\
}

\begin{document}

\maketitle

\begin{abstract}
	The interpretation of the origin of observed exoplanets is usually done only qualitatively due to uncertainties of key parameters in planet formation models. To allow a quantitative methodology which traces back in time to the planet birth locations, we train recently developed conditional invertible neural networks (cINN) on synthetic data from a global planet formation model which tracks growth from dust grains to evolved final giant planets. In addition to deterministic single planet formation runs, we also include gravitationally interacting planets in multiplanetary systems, which include some measure of chaos. For the latter case, we treat them as individual planets or choose the two or three planets most likely to be discovered by telescopes. We find that training on multiplanetary data, each planet treated as individual point, is promising. The single-planet data only covers a small range of planets and does not extrapolate well to planet properties not included in the training data. Extension to planetary systems will require more training data due to the higher dimensionality of the problem. 
\end{abstract}

\section{Motivation}
Planet formation theory has advanced thanks to the discovery of thousands of exoplanets. In an optimistic view, successful planet formation models exist and can reproduce the general outline of the exoplanet demographics. However, some planet formation parameters are notoriously difficult to constrain, such as a parameter characterizing turbulence in the disk, while others are known to vary significantly within a certain range, such as the initial disk gas and solid mass. It is therefore interesting to try to constrain these values by retrieving planet formation parameters from the exoplanet data \citep{Molliere2022}. Therefore, we attempt here a first step towards this goal, by starting with planetary mass and distance to the star as observables. A direct MCMC approach is not feasible due to the long computation time of the physical model (days to months for the setup used here, although a simple parametric model has been used by \citealp{Chambers2018}). Therefore, a surrogate model would be required in an MCMC search. If training of a surrogate model is required in any case, we want to assess here whether an invertible neural network can be used. The overarching goal is to apply the model to the exoplanet data to retrieve the likelihood of formation parameters given a physical model as well as to improve the quantitative interpretation of individual systems.

\section{Methods}

\subsection{Physical model}
To study planet formation from a global perspective, we use an established global model \citep[described in detail by][]{Emsenhuber2020a} with extensions from \citet{Voelkel2020,Voelkel2022}. In the following, we assume that the model is physically correct and includes all relevant ingredients. This is a strong assumption and neglects for example the effects of continous infall of matter and assumes that viscosity and not magnetic effects drive the observed accretion of gas by the star. For the sake of a methodological investigation, we assume the model is true.

Here, a brief summary follows. At an early stage, a relatively massive disk has formed with a mass fraction $M_{\rm disk}/M_{\star}$ relative to the central star. It is mostly gaseous, with a percent-level dust content, that is, the solid disk mass $M_{\rm solid, disk}$, set by the dust-to-gas ratio. Its initial radius follows an empirical relation for an early disk stage \citep{Tobin2020}: $R_{\rm disk} = 70\,{\rm au} \times \left(M_{\rm solid, disk}/(100 M_{\oplus})\right)^{1/4}$. Another varied initial condition, is the disk inner edge defined by magnetospheric accretion onto the star.

A central parameter controlling several aspects of planet formation is the non-dimensional viscous $\alpha$ parameter \citep{Shakura1973}, assumed to arise from turbulence but still poorly constrained. By assuming a value for $\alpha$, the model evolves the one-dimensional viscous diffusion equation numerically \citep{Pringle1981}. The parameter also influences disk temperatures and planetary migration. 
	
The model evolves the solid material by balancing coagulation and fragmentation as well as radial drift driven by aerodynamic breaking of the orbiting particles \citep{Birnstiel2012}. Gas turbulence sets the relative speed of particles and thus whether they fragment. For both this process and vertical dust settling, we use a reduced value of $\alpha_{\rm dust} = 0.1\alpha$, not physically motivated but chosen to allow sufficient pebble accretion, which would otherwise be suppressed. It is however conceivable that turbulence on small scales and in vertical direction differs from the global, radial one \citep{Lesur2023}. 

Radial drift implies a mass flux of solids, mainly centimeter-sized \textit{pebbles}, from which a fraction of $\epsilon_{\rm plts}$ (set to 0.01) is converted into larger bodies called \textit{planetesimals} over a characteristic length scale of five disk scale heights \citep{Lenz2019}. Theory and Solar System evidence suggest planetesimals are at least 50\,km in diameter \citep{Polak2023}, which we assumed here. The planetesimal size is important for planetesimal collision rates, which we treat in two stages. In the early runaway regime, we assume that a largest planetesimal -- given a head-start with a larger initial diameter of 1320\,km -- accretes smaller planetesimals and pebbles \citep[following][]{Ormel2017a}. Once it reaches 0.01\,Earth masses, the body is promoted to a protoplanet in the disk model, from which point onward it influences planetesimal dynamics \citep{Fortier2013}, removes mass from the solid and gas disk via accretion, and interacts with other protoplanets gravitationally which is a source of chaos \citep[using the symplectic N-body code by][]{Chambers1999}. This two-stage process reduces computation for bodies with negligible mass.

The protoplanets can also accrete gas, where the rate is obtained from solving the one-dimensional interior structure of the gaseous envelope and limited by accretion rates obtained from detailed three-dimensional calculations \citep{Bodenheimer2013}. Ultimately, gas accretion ceases when the disk dissipates after several million years through so-called \textit{photoevaporation}, thermal mass loss driven by heation from stellar X-ray \citep{Picogna2019,Ercolano2021} and external ultraviolet irradiation \citep{Haworth2018}, with parameters listed in Table \ref{tab:parameters}.

\subsection{Data}
Two datasets are generated with this set-up, one where at most one protoplanet forms and injection of further protoplanets is suppressed (single-planet case), and one which allows up to 100 protoplanets per disk (nominal). The initial conditions are varied in logarithmic space in the four dimensions listed in Table \ref{tab:parameters}.  For both cases, 1000 simulations were started. In some disks, the conditions did not allow a single protoplanet to grow to the threshold mass which results in 707 planets for the single-planet data and 15777 planets in 690 disks for the multi-planet case, where some simulations did not successfully complete due to known numerical issues. From the multi-planet case, we also extracted the two and three planets with the highest radial velocity signal imposed on their stars ordered by this radial velocity semi-amplitude. This gives us two additional datasets with 679 (two-planet) and 658 (three-planet) entries. We note that for training an invertible neural network with four input dimensions, these datasets are relatively small which makes the task challenging and further data generation is conceivable in the future if motivated by the early results presented below. For slight data augmentation, we re-draw noise from a Gaussian model with deviation of 0.01 in normalized units for both parameters (disk mass fraction, $\alpha$, dust/gas ratio, and inner edge, all logarithmic) and observations (the planets' mass and semi-major axis, also logarithmic) at each training epoch.

\subsection{Neural network design and training}
\label{sec:ML_model}
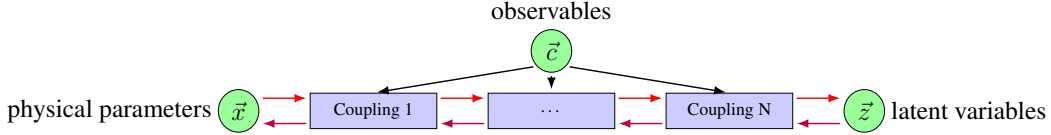
\begin{figure}
	\centering
	\resizebox{\linewidth}{!}{%
		\begin{tikzpicture}[
			block/.style={rectangle, draw, minimum width=2.5cm, minimum height=0.7cm, align=center, fill=blue!20,font=\normalfont},
			data/.style={draw, ellipse, minimum size=0.8cm, align=center,fill=green!40, font=\Large},
			fflowarrow/.style={-{Latex}, thick, shorten >=2pt, shorten <=2pt,font=\normalsize, red},
			iflowarrow/.style={-{Latex}, thick, shorten >=2pt, shorten <=2pt,font=\normalsize, purple},
			cflowarrow/.style={-{Latex}, thick, shorten >=2pt, shorten <=2pt,font=\normalsize},
			node distance=1.6cm and 1.cm
			]
			
			\node[data] (x) {$\vec{x}$};
			\node[block, right=of x] (ac1) {Coupling 1};
			\node[block, right=of ac1] (ac2) {\dots};
			\node[block, right=of ac2] (acN) {Coupling N};
			\node[data, right=of acN] (z) {$\vec{z}$};
			\node[data, above=0.4cm of ac2] (c) {$\vec{c}$};
			
			\node[font=\Large, anchor=east, align=right]  at (x.west)  {physical parameters};
			\node[font=\Large, anchor=west, align=left] at (z.east) {latent variables};
			\node[font=\Large, above=2pt of c] {observables};
			
			\draw[fflowarrow] ($(x.east)+(0,0.25cm)$) -- ($(ac1.west)+(0,0.25cm)$);
			\draw[fflowarrow] ($(ac1.east)+(0,0.25cm)$) -- ($(ac2.west)+(0,0.25cm)$);
			\draw[fflowarrow] ($(ac2.east)+(0,0.25cm)$) -- ($(acN.west)+(0,0.25cm)$);
			\draw[fflowarrow] ($(acN.east)+(0,0.25cm)$) -- ($(z.west)+(0,0.25cm)$);
			
			\draw[iflowarrow] ($(ac1.west)+(0,-0.25cm)$) -- ($(x.east)+(0,-0.25cm)$);
			\draw[iflowarrow] ($(ac2.west)+(0,-0.25cm)$) -- ($(ac1.east)+(0,-0.25cm)$);
			\draw[iflowarrow] ($(acN.west)+(0,-0.25cm)$) -- ($(ac2.east)+(0,-0.25cm)$);
			\draw[iflowarrow] ($(z.west)+(0,-0.25cm)$) -- ($(acN.east)+(0,-0.25cm)$);
			
			\draw[cflowarrow] (c.south west) -- (ac1.north);
			\draw[cflowarrow] (c.south) -- (ac2.north);
			\draw[cflowarrow] (c.south east) -- (acN.north);
			
		\end{tikzpicture}%
	}
	\caption{cINN information flow. Red: forward direction; Purple: inverse; Black: conditions (required for both). Observables $\vec{c}$, planetary mass and semi-major axis, are from the training, test, or real data; model parameters $\vec{x}$ are here $M_{\rm disk}$, $\alpha$, inner edge orbital period, and dust-to-gas ratio.}
	\label{fig:network_overview}
\end{figure}

In this work, we make use of conditional invertible neural networks (cINN) \citep{Ardizzone2019,Ardizzone2021} by extending the public Framework for Easily Invertible Architectures (FrEIA: \hyperref{https://github.com/vislearn/FrEIA}{}{}{https://github.com/vislearn/FrEIA}, \citealp{freia}). The structure of cINN deviates from invertible neural networks (INN) presented in \citet{Ardizzone2018} in that the training or real data is not part of the mapping of parameters to observables of the INN, but instead, it is given to all conditional blocks in the network structure. Figure \ref{fig:network_overview} schematically depicts the approach. A commodity of both approaches is that the model is trained to map parameters ($\vec{x}$) to a latent variable space $\vec{z}$ without physical meaning but with a distribution in $\vec{z}$ space following a multidimensional Gaussian with identity covariance matrix. Upon successful training, this property can be used to draw $\vec{z}$ values from the Gaussian, transverse the network in inverse direction under conditions of true observed systems and infer physical parameters.

The detailed structure of the network trained here follows \citet{Ksoll2020} based on GLOW-style \citep{Kingma2018} affine coupling layers with conditioning. Compared to \citet{Ksoll2020}, we reduced the network width and optimized training parameters for our data. We use 16 coupling blocks with random permutations between them, each containing 3 hidden layers with 8 units per layer and rectified linear unit activation, motivated by our low-dimensional $\vec{c}$ and $\vec{x}$. Future iterations should explore the hyperparameter range in further depth to optimize it for efficiency. Training is performed by minimizing with the Adam optimizer ($\beta_1=0.8$ and $\beta_2=0.8$, initial learning rate of 0.001, decaying using the StepLR function with $\gamma= 0.99$ and step sizes of one epoch) the negative log likelihood from the forward passes $f(\vec{x};\vec{c})$ as well as the mean squared errors between recovered parameters $\hat{\vec{x}}$ and true parameters $\vec{x}$, resulting in the total loss:
\[
\mathcal{L} = 
\frac{1}{2} \| f(\vec{x}; \vec{c}) \|^2 
- \log \Big| \det \frac{\partial f(\vec{x}; \vec{c})}{\partial \vec{x}} \Big|
+ \, \| \hat{\vec{x}} - \vec{x} \|^2\,.
\]
For the multiplanet sample treated as individual planets, a validation set of 10 batches of 64 points is visualized during training and learning parameters were tuned manually using this measure. A test set of 1577 points (10\%) is held out from training and parameter tuning completely until testing of the results.

\section{Results}
\begin{figure}
	\centering
	\includegraphics[width=0.3\linewidth]{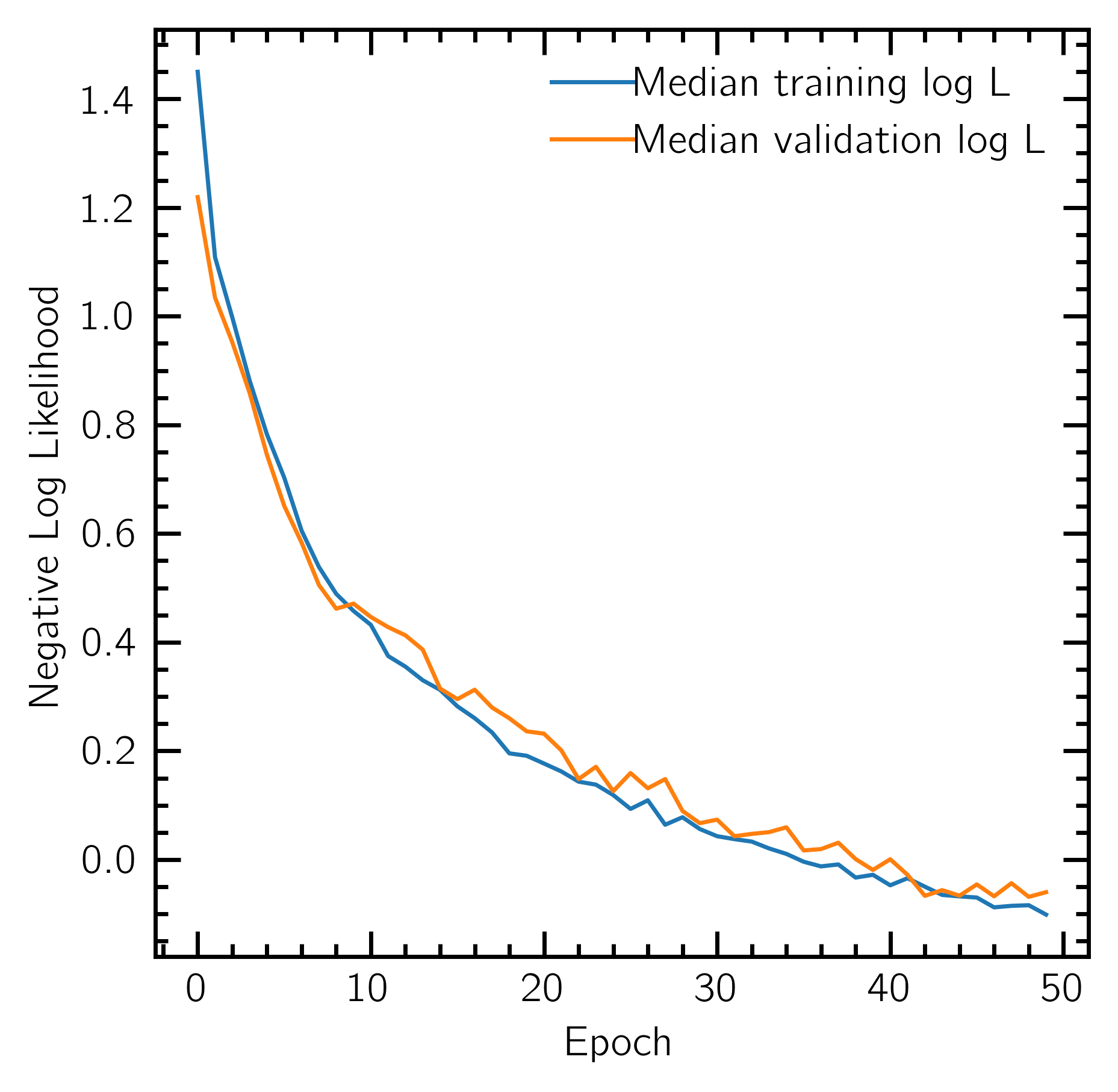}
	\includegraphics[width=.33\linewidth]{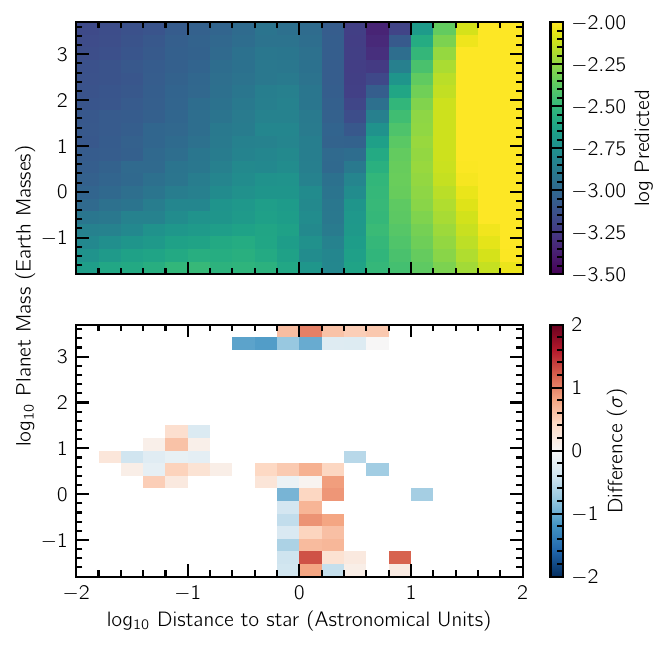}
	\includegraphics[width=.33\linewidth]{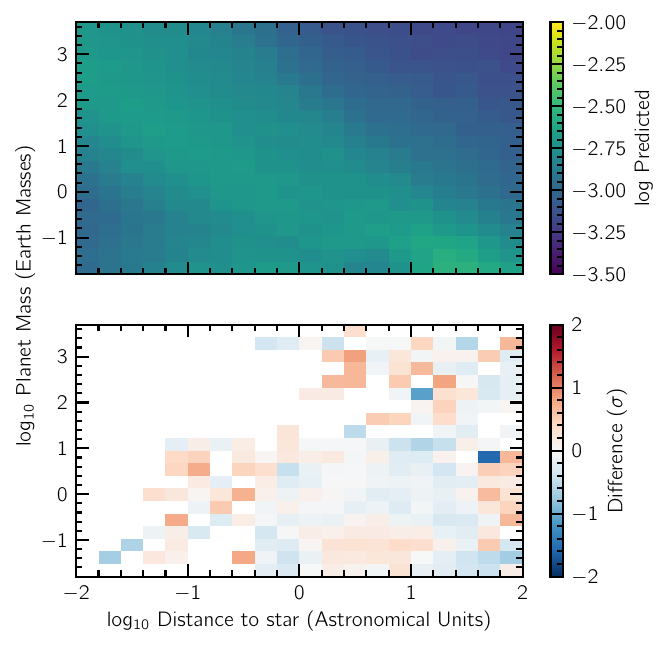}
	\caption{Loss function during training and comparison of predicted and training data on the viscous $\alpha$. Left: Validation and training Log Likelihood as a function of training epoch. Center: single-planet histogram on inferred $\alpha$; Right: as Center but for the nominal data. Difference of the posterior mean to training data is normalized by standard deviations as  $\sigma = \sqrt{\sigma_{\rm train}^2 + \sigma_{\rm posteriors}^2}$.}
	\label{fig:training_acc}
\end{figure}

\begin{figure}
    \centering
    \textbf{\qquad\qquad\quad Nominal \qquad\qquad\qquad Single \qquad\qquad 2 detectable planets}\\
    \includegraphics[width=0.23\linewidth]{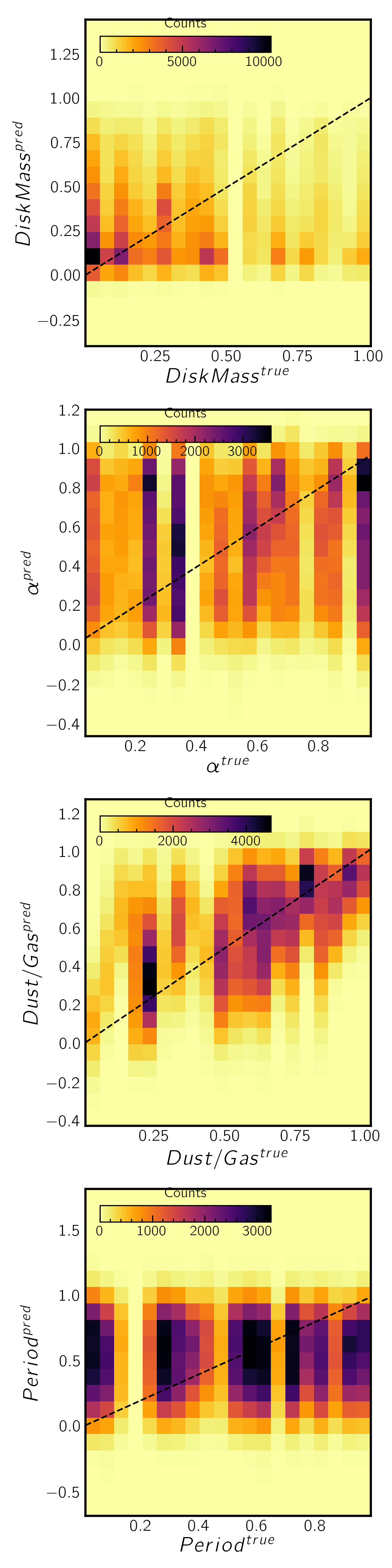}
    \includegraphics[width=0.23\linewidth]{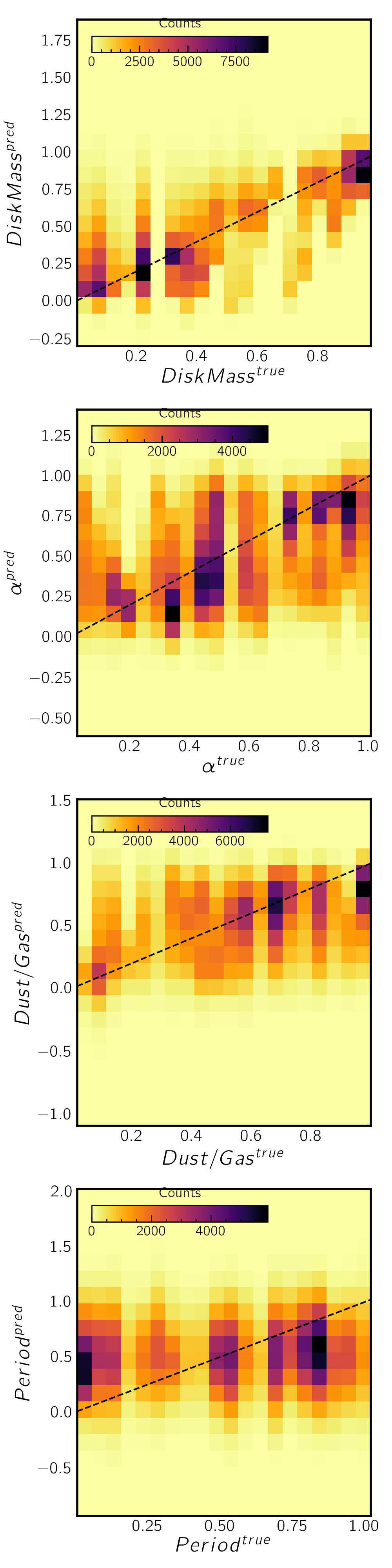}
    \includegraphics[width=0.23\linewidth]{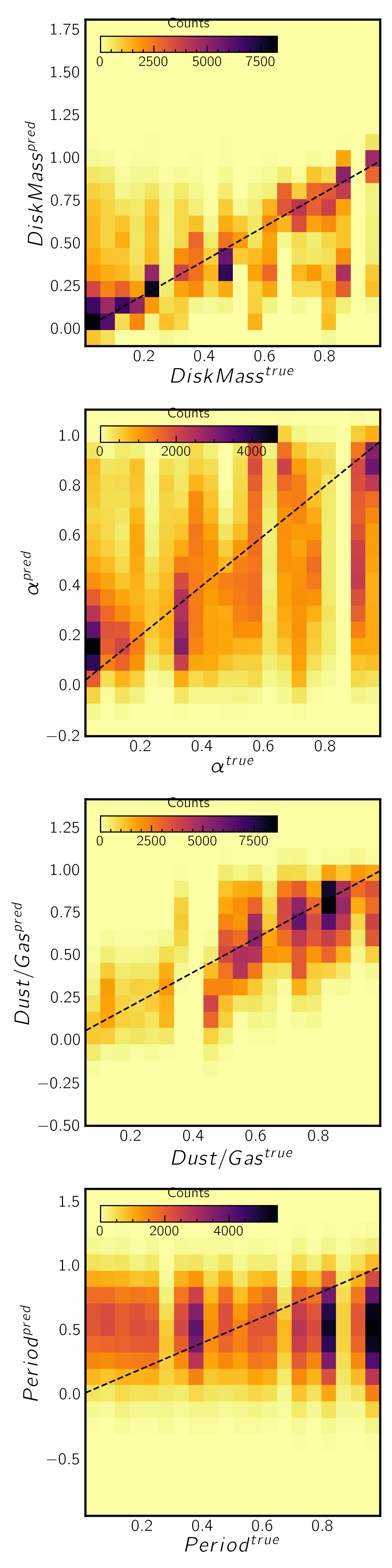}
    \caption{True values against posterior distributions in normalized units for the nominal, single-planet, and 2 detectable planet cases.}
    \label{fig:true_vs_posteriors}
\end{figure}

Visual inspection reveals that the loss function reaches its optimum after about 50 epochs, at which point we stopped training (left panel in Fig. \ref{fig:training_acc}). Accuracy is measured in several ways. First, maximum a posteriori estimates (MAP) are compared with the ground truth data from the test set with satisfying results. For example, for the nominal data, we find average deviations of 0.2 in standardized units, centered on zero. For all datasets, both the training data and the cINN results are insensitive (within the achieved accuracy) to the inner-edge orbital period; we therefore omit this parameter from the following discussion. We further tested posterior distributions agains the true values (Fig. \ref{fig:true_vs_posteriors}) with better matches for the non-single planet cases and the scatter around the true value indicating that with limited data (e.g. a single planet), no strong inference of the four disk parameters can be made.

For the single-planet data, the observation parameter space, from which $\vec{c}$ are drawn, is poorly sampled, with typical planet formation outcomes clustering in restricted regions. The cINN therefore extrapolates to unsampled areas, as seen in the central panel of Figure \ref{fig:training_acc}: performance is reasonable where data exists, but at long orbital distances the model predicts large turbulence parameters. Such a dependency is not physically expected and the results from multiplanet (nominal) data, which better sample this space, does not show such behavior. More concerning than the extrapolation itself is that the posterior width narrows in this region. The model therefore predicts with confidence an unexpected and unwarranted extrapolation which is concerning in applications. Although this effect is weaker in other parameters, we conclude that single-planet simulations are unsuitable for building a robust model applicable to real exoplanet data which includes exoplanets at these locations.

The nominal data shows a more expected dependency on $\alpha$. The cINN predicts a diagonal of enhanced values in distance against mass space (Fig. \ref{fig:training_acc}, right), which we attribute to the interplay of the effects of $\alpha$ on both migration (influencing more massive planets) and dust properties (determining the low-mass growth).

\begin{figure}
	\centering
	\includegraphics[width=.329\linewidth]{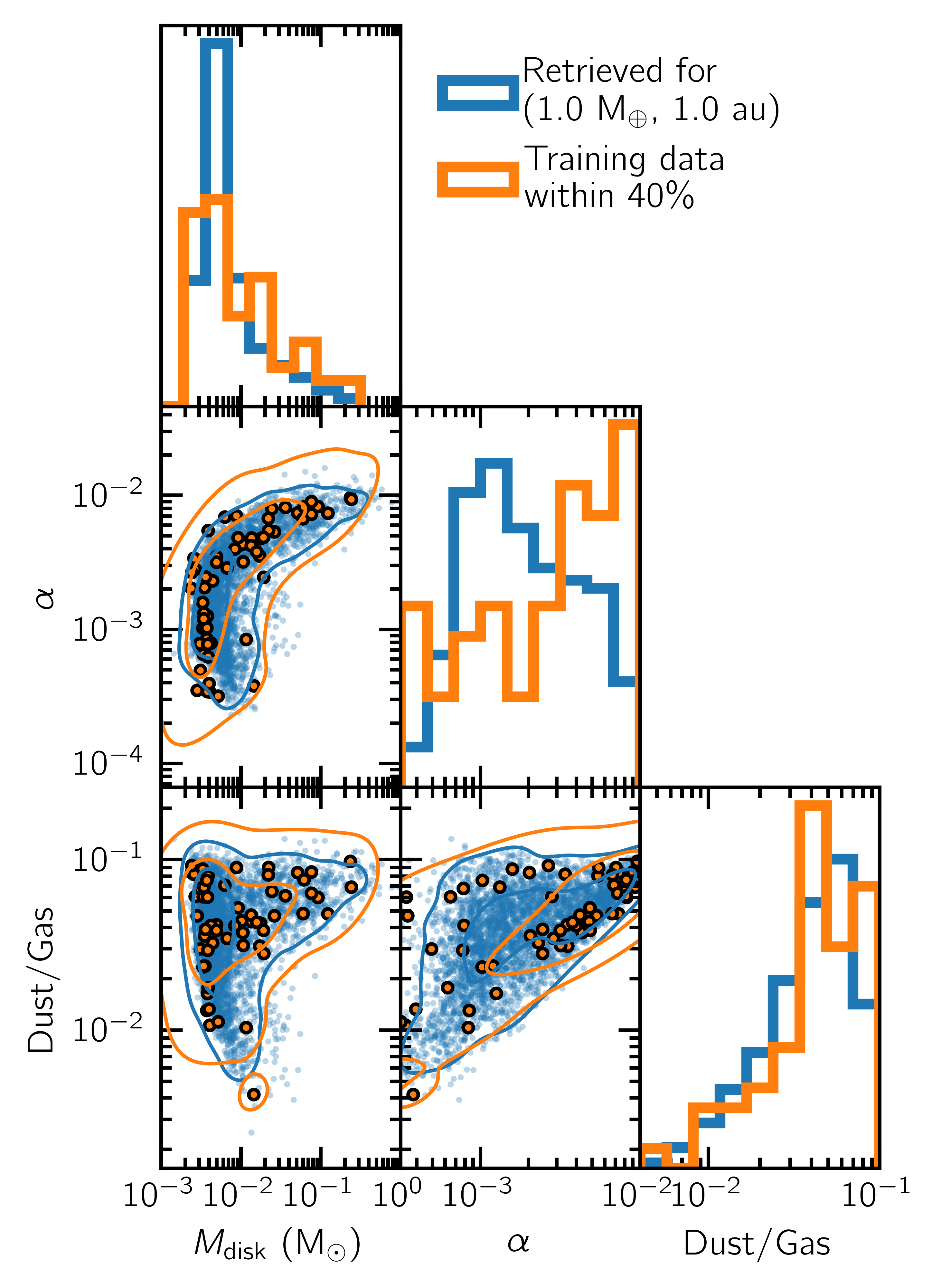}
	\includegraphics[width=.329\linewidth]{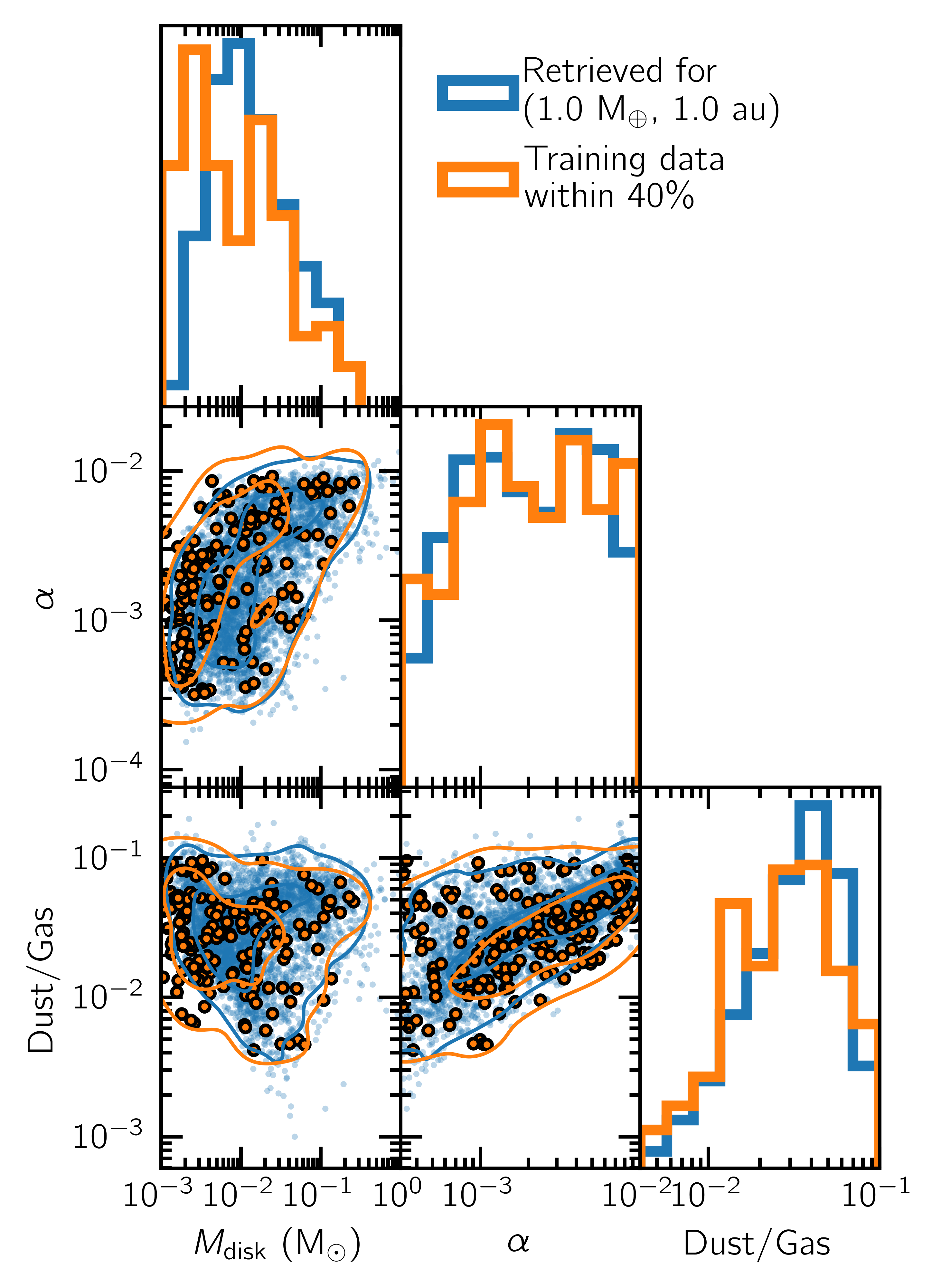}
	\includegraphics[width=.329\linewidth]{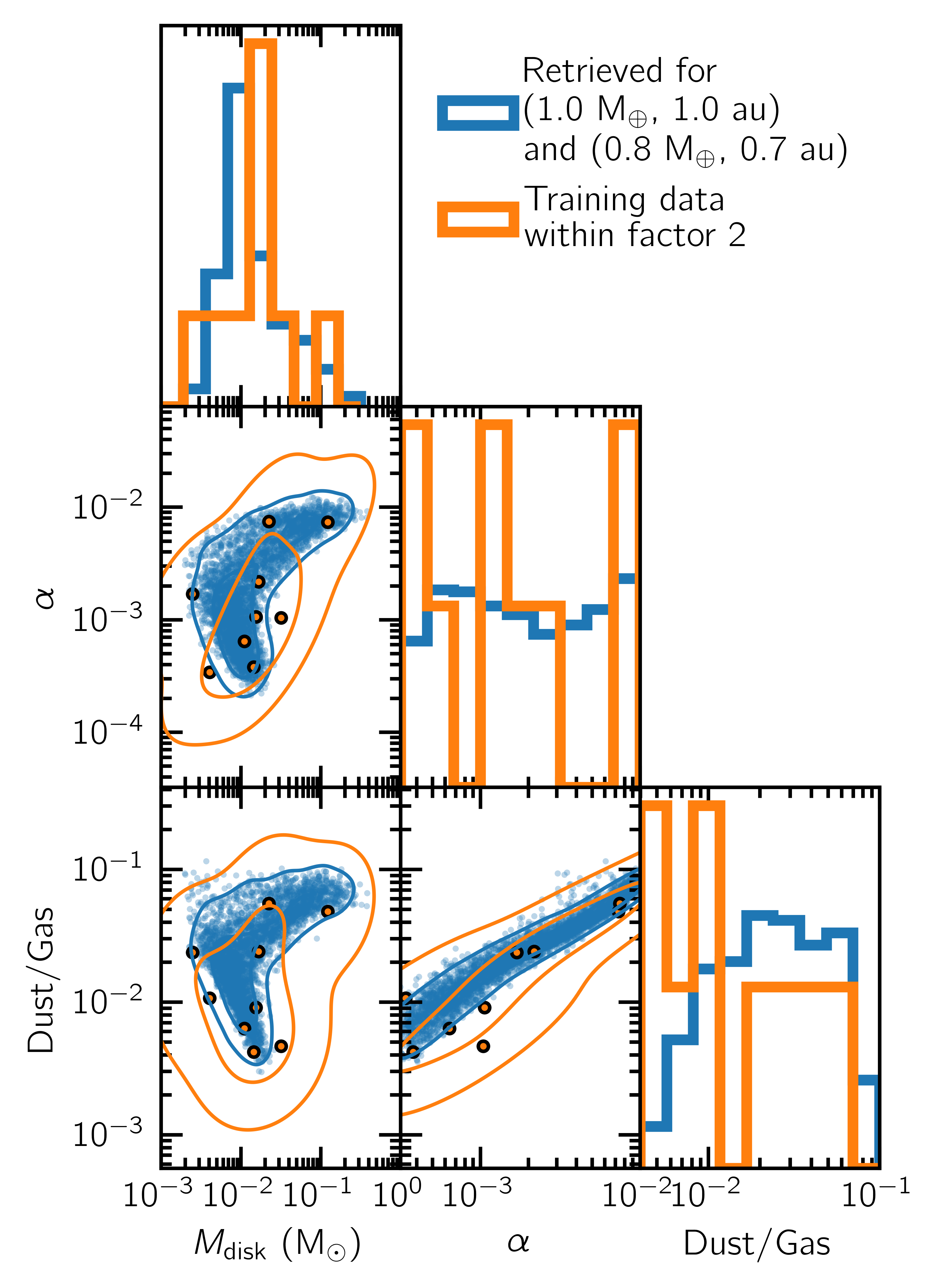}
	\caption{Retrieved and training disk parameters for Earth-analogues. Left: single-planet per disk; Center: nominal, Right: two-planet systems. Training data is shown if both the planet mass and semi-major axis lies within 40\% of the Earth (factor 2 of an Earth + Venus system).}
	\label{fig:scatter}
\end{figure}

As a first application and further test, we let the models predict the disk parameters for selected planets and contrast them against the full test and training data in corner plots (e.g. for an Earth-like planet in Fig. \ref{fig:scatter}). The retrievals appear qualitatively reasonable and reveal correlations between important parameters, $M_{\rm disk}$, $\alpha$, and Dust/Gas ratio, accross all tested observation $\vec{c}$ combinations (0.1, 1, 10, and 4000 Earth mass planets at 0.1, 1, or 10\,au). Comparisons with training data are limited to $\vec{c}$ regions with sufficient samples. For the two-planet systems (four dimensional $\vec{c}$), only a few hundred test cases are available, leading to sparse coverage and necessitating a broader range of included training data for the plot. The issue becomes more severe for three-planet systems (six-dimensional). Nevertheless, training seems to converge to an optimum, and the MAP estimates remain similarly close to the test samples as in the two cases with individual planets (single-planet and nominal). We acknowledge that more rigorous testing is required before the approach can be applied.

In summary, we conclude that a cINN can be trained on the outcomes of a global dust-to-planet formation model. Multiplanet simulations sample more evenly the parameter space and yield improved results and applicability. In these simulations, the introduced chaos is not detrimental and might even lead to a more robust training, indicating that the imprint of disk parameters survives planet-planet interactions. To accommodate planetary system data, focusing on the most observable planets is practical, and data generation efficiency could be improved by raising the mass threshold at which protoplanets are individually resolved. Future work is motivated by this study which should rely on a larger set of training data and compare the novel cINN approach against a traditional MCMC approach with a forward network as surrogate.

\begin{ack}
	R.B. acknowledges the financial support from DFG under Germany’s Excellence Strategy EXC 2181/1-390900948, Exploratory project EP 8.4 (the Heidelberg STRUCTURES Excellence Cluster).
	The work has made extensive use of the pytorch \citep{pytorch} package.
	The plots shown in this work were generated using \textit{matplotlib} \citep{Hunter2007} and seaborn \citep{seaborn}.
\end{ack}

\bibliographystyle{aa} 
\bibliography{library_betterbib_zotero} 

\appendix

\section{Technical Appendices and Supplementary Material}
We provide a table of physically relevant parameters for the generation of the dataset in Table \ref{tab:parameters}. These parameters are required for reproducibility of our results since they partially deviate from prior published works. Furthermore, we provide here further insight on the quality of the training by providing normalized posterior distributions against true values.

\begin{table}[H]
	\caption{Key physical model parameters}
	\label{tab:parameters}
	\centering
	\begin{tabular}{lll}
		\toprule
		\multicolumn{3}{c}{Varied parameters (uniform sampling)}        \\
		\midrule
		$\log_{10} M_{\rm disk}/M_{\star}$     & Gas disk mass fraction     & [-3,-0.5] \\
		$\log_{10} \alpha$     & Viscous parameter     & [-3.5,-2] \\
		$\log_{10} M_{\rm dust}/M_{\rm gas}$ & Dust/Gas ratio & [-2.4,-1] \\
		$\log_{10} P_{\rm in}$ & Inner edge orbital period & [0, 1.3]\hyperref[ftn:1]{$^*$}\\
		\midrule
		\multicolumn{3}{c}{Constants}        \\
		\midrule
		$N_{\rm p,max}$ & Maximum number of planets & [1,100]\\
		$M_{\star}$ & Stellar mass & 1\,Solar mass\\
		$R_{\rm plts}$ & Planetesimal radius & 25\,km    \\
		$R_{\rm l,plt}$ & Largest planetesimal radius & 660\,km    \\
		$v_{\rm frag}$ & Pebble fragmentation velocity & 1\,m/s\\
		$L_{\rm X}$ & X-ray luminosity for disk photoevaporation & $5\times 10^{28}$ erg/s\\
		$F_{UV}$ & External ultraviolet field strength & 10\,G$_0$\hyperref[ftn:2]{$^\dagger$}\\
		$\epsilon_{\rm plts}$ & Planetesimal formation efficiency & 0.01\\
		$T_{\rm min}$ & Minimum disk temperature & 10\,K\\
		$f_{\rm opa}$ & Opacity reduction factor in planet envelope & 0.003\hyperref[ftn:3]{$^\ddagger$}\\
		$\rho_{\rm dust}$ & Dust bulk density & 1.675\,g/cm$^3$\\
		$a_0$ & Dust monomer (minimum) size & $10^{-5}$\,cm\\
		\bottomrule
		\multicolumn{3}{l}{\footnotesize \label{ftn:1}$^*$Corresponds to a range from 1 to 20 days}\\
		\multicolumn{3}{l}{\footnotesize \label{ftn:2}$^\dagger$Measured in local interstellar FUV  (912-2400\,\AA) radiation field strength G$_0 = 1.6\times 10^{-3}$\,erg/(s cm$^{2}$)}\\
		\multicolumn{3}{l}{\footnotesize \label{ftn:3}$^\ddagger$Following the dust growth argument by \citet{Mordasini2014}}
	\end{tabular}
\end{table}

\end{document}